\documentclass
[aps,twocolumn,noshowpacs,noshowkeys,letterpaper,superscriptaddress]{revtex4}%
\usepackage{amsfonts}
\usepackage{amsmath}
\usepackage{amssymb}
\usepackage{graphicx}
\usepackage{textcomp}%
\begin{document}
\title{Optical frequency comb generation from a monolithic microresonator}
\author{P.\ Del'Haye}
\affiliation{Max Planck Institut f\"ur Quantenoptik, 85748 Garching, Germany}
\author{A.\ Schliesser}
\affiliation{Max Planck Institut f\"ur Quantenoptik, 85748 Garching, Germany}
\author{O.\ Arcizet}
\affiliation{Max Planck Institut f\"ur Quantenoptik, 85748 Garching, Germany}
\author{T.\ Wilken}
\affiliation{Max Planck Institut f\"ur Quantenoptik, 85748 Garching, Germany}
\author{R.\ Holzwarth}
\affiliation{Max Planck Institut f\"ur Quantenoptik, 85748 Garching, Germany}
\author{T.J.\ Kippenberg}
\email{tjk@mpq.mpg.de}
\affiliation{Max Planck Institut f\"ur Quantenoptik, 85748 Garching, Germany}
\homepage{www.mpq.mpg.de/k-lab/}
\keywords{Cooling, opto-mechanical coupling, radiation pressure, micro-mechanical
oscillator, dynamical backaction}
\pacs{PACS number: 42.65.Sf, 42.50.Vk}

\begin{abstract}
Optical frequency combs\cite{Udem2002, Cundiff2003, BookFemto2005} provide
equidistant frequency markers in the infrared, visible and
ultra-violet\cite{Jones2005, Gohle2005} and can link an unknown optical
frequency to a radio or microwave frequency reference\cite{Diddams2000,
Diddams2001}. Since their inception frequency combs have triggered major
advances in optical frequency metrology and precision
measurements\cite{Diddams2000, Diddams2001} and in applications such as
broadband laser-based gas sensing\cite{Thorpe2006} and molecular
fingerprinting\cite{Diddams2007}. Early work generated frequency combs by
intra-cavity phase modulation\cite{KOUROGI1993, Ye1997}, while to date
frequency combs are generated utilizing the comb-like mode structure of
mode-locked lasers, whose repetition rate and carrier envelope phase can be
stabilized\cite{Jones2000a}. Here, we report an entirely novel approach in
which equally spaced frequency markers are generated from a continuous wave
(CW) pump laser of a known frequency interacting with the modes of a
monolithic high-Q microresonator\cite{Armani2003} via the Kerr
nonlinearity\cite{Kippenberg2004a, Savchenkov2004a}. The intrinsically
broadband nature of parametric gain enables the generation of discrete comb
modes over a 500 nm wide span ($\approx70$ THz) around 1550 nm without relying
on any external spectral broadening. Optical-heterodyne-based measurements
reveal that cascaded parametric interactions give rise to an optical frequency
comb, overcoming passive cavity dispersion. The uniformity of the mode spacing
has been verified to within a relative experimental precision of
$7.3\times10^{-18}$. In contrast to femtosecond mode-locked
lasers\cite{Steinmeyer1999} the present work represents an enabling step
towards a monolithic optical frequency comb generator allowing significant
reduction in size, cost and power consumption. Moreover, the present approach
can operate at previously unattainable repetition rates\cite{Keller2003}
exceeding 100 GHz which are useful in applications where the access to
individual comb modes is required, such as optical waveform
synthesis\cite{Weiner2000}, high capacity telecommunications or astrophysical
spectrometer calibration\cite{murphyar2007}. \newline

\end{abstract}
\maketitle

Optical microcavities\cite{Vahala2003} are owing to their long temporal and
small spatial light confinement ideally suited for nonlinear frequency
conversion, which has led to a dramatic improvement in the threshold of
nonlinear optical light conversion\cite{Bookchang}. In contrast to stimulated
gain, parametric frequency conversion\cite{Dunn1999} does not involve coupling
to a dissipative reservoir, is broadband as it does not rely on atomic or
molecular resonances and constitutes a phase sensitive amplification process,
making it uniquely suited for tunable frequency conversion. In the case of a
material with inversion symmetry - such as silica - the non linear optical
effects are dominated by the third order non linearity. The process is based
on four-wave mixing among two pump photons (frequency $\nu_{P}$) with a signal
($\nu_{S}$) and idler photon ($\nu_{I}$) and results in the emergence of
(phase coherent) signal and idler sidebands from the vacuum fluctuations at
the expense of the pump field (cf. Fig.1). The observation of parametric
interactions requires two conditions to be satisfied. First momentum
conservation has to be obeyed. This is intrinsically the case in a whispering
gallery type microcavity\cite{Vahala2003} since the optical modes are angular
momentum eigenstates and have (discrete) propagation constants $\beta
_{m}=\frac{m}{R}$ resulting from the periodic boundary condition, where the
integer $m$ designates the mode number and $R$ denotes the cavity radius.
Hence the conversion of two pump photons (propagation constant $\beta_{N}$)
into adjacent signal and idler modes ($\beta_{N-\Delta N}$, $\beta_{N+\Delta
N}$, $\Delta N=1,2,3\ldots$) conserves momentum
intrinsically\cite{Kippenberg2004a} (analogous reasoning applies in the case
where the two annihilated photons are in different modes, i.e. for four-wave
mixing, cf. Fig. 1b). The second condition that has to be met is energy
conservation. As the parametric process creates symmetrical sidebands with
respect to the pump frequency (obeying $h\nu_{I}+h\nu_{S}=2h\nu_{P}$, where
$h$ is the Planck constant) it places stringent conditions on the cavity
dispersion that can be tolerated since it requires a triply resonant cavity.
This is a priori not expected to be satisfied, since the distance between
adjacent modes $\nu_{\mathrm{FSR}}=\left\vert \nu_{m}-\nu_{m+1}\right\vert $
(the free spectral range, FSR) can vary due to both material and intrinsic
cavity dispersion which impact $n_{\mathrm{eff}}$ and thereby render optical
modes (having frequencies $\nu_{m}=m\cdot\frac{c}{2\pi\cdot R\cdot
n_{\mathrm{eff}}}$, where $c$ is the speed of light in vacuo) non-equidistant.
Indeed, it has only recently been possible to observe these processes in
microcavities (made of crystalline\cite{Savchenkov2004a} CaF$_{2}$ and
silica\cite{Kippenberg2004a, Carmon2007}). \newline Importantly, this
mechanism could also be employed to generate optical frequency combs: the
initially generated signal and idler sidebands can interact among each other
and produce higher order sidebands (cf. Fig. 1) by non-degenerate four-wave
mixing (FWM)\cite{STOLEN1982} which ensures that the frequency difference of
pump and first order sidebands $\Delta\nu\equiv\left\vert \nu_{P}-\nu
_{S}\right\vert =\left\vert \nu_{P}-\nu_{I}\right\vert $ is \textit{exactly}
transferred to all higher order sidebands. This leads to an equidistant
spectrum throughout the \textit{entire} comb. This can be readily seen by
noting that e.g. the 2$^{\mathrm{nd}}$ order sidebands are generated by mixing
among the pump and first order signal/idler sidebands (e.g. $\nu_{S2}=\nu
_{P}+\nu_{S}-\nu_{I}=\nu_{P}-2\Delta\nu$), which rigidly determines the
spacing of any successively higher sidebands. Thus, provided the cavity
exhibits low dispersion, the successive four-wave mixing to higher orders
would intrinsically lead to the generation of phase coherent sidebands with
equal spacing, \textit{i.e. an optical frequency comb}. Here, we report that
microresonators allow realization of this process and generation of optical
frequency combs.

\begin{figure}[ptb]
\begin{center}
\includegraphics[width=7.5cm]{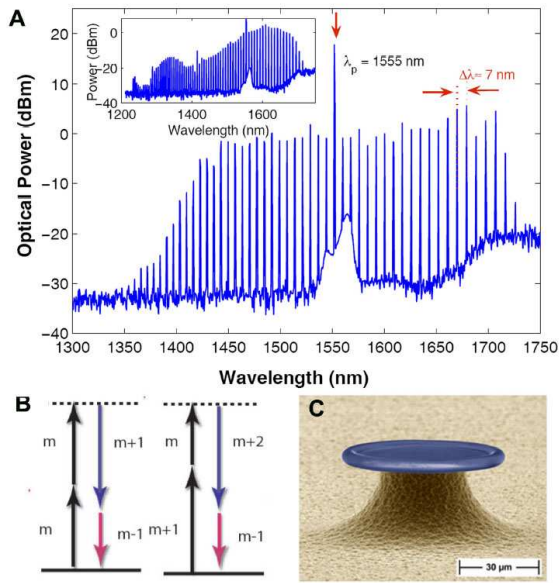}
\end{center}
\caption{{}Broadband parametric frequency conversion from a 75-$\mu
m$-diameter monolithic microresonator. Panel A: Spectrum of the parametric
frequency conversion observed in an 75-$\mu m$-diameter monolithic toroid
microcavity when pumped with 60 mW continuous wave (CW) laser power at 1555
nm. The combination of parametric interactions and four-wave-mixing (FWM)
gives rise to a broadband emission, spaced by the cavity free spectral range.
Inset: Broadband parametric conversion of a different sample generating more
than 70 parametric modes extending over a wavelength span of nearly 500 nm
(launched power 130 mW). The asymmetry in the spectrum (with higher power in
the higher wavelength sidebands) and the amplitude modulation of the emission
is attributed to Raman amplification and variation of the taper fiber output
coupling, respectively. Panel (B): Schematic of the processes that contribute
to the parametric conversion: degenerate (left) and non-degenerate (right)
four-wave-mixing among cavity eigenmodes. Momentum conservation is
intrinsically satisfied for the designated angular mode number (m)
combinations. Panel (C): Scanning electron microscope image of a toroid
microcavity on a silicon chip. }%
\end{figure}We employ ultra-high-Q monolithic microresonators in the form of
silica toroidal microcavities\cite{Armani2003} on a silicon chip, which
possess giant photon storage times ($\tau$) i.e. ultra-high quality factors
($Q=2\pi\nu\tau>10^{8}$) and small mode volumes. Highly efficient coupling is
achieved using tapered optical fibers\cite{Spillane2003}. Owing to the high
circulating power, parametric interactions are readily observed at a threshold
of approx. 50 $\mu$W. When pumping with a continuous wave (CW) 1550-nm laser
source, we observe a massive cascade and multiplication of the parametric
sidebands extending to both higher and lower frequencies. Fig. 1a shows a
spectrum in which a 75-$\mu$m-diameter microcavity was pumped with 60 mW
power, giving rise to an intra-cavity intensity exceeding 1 GW/cm$^{2}$. The
parametric frequency conversion could extend over more than 490 nm (cf. Fig 1a
inset), with the total conversion efficiency being 21.2 \% (The highest
observed conversion efficiency was 83 \% by working in the over-coupled
regime\cite{Armani2003}). These bright sidebands (termed Kerr combs in the
remaining discu
ssions) could be observed in many different samples. Also, in
the largest fabricated samples (190 $\mu$m diameter) 380-nm broad Kerr combs
comprising 134 modes spaced by 375 GHz could be generated at the expense of
slightly higher pump power (cf. Supplementary Information).

\begin{figure}[tbh]
\begin{center}
\includegraphics[width=8cm]{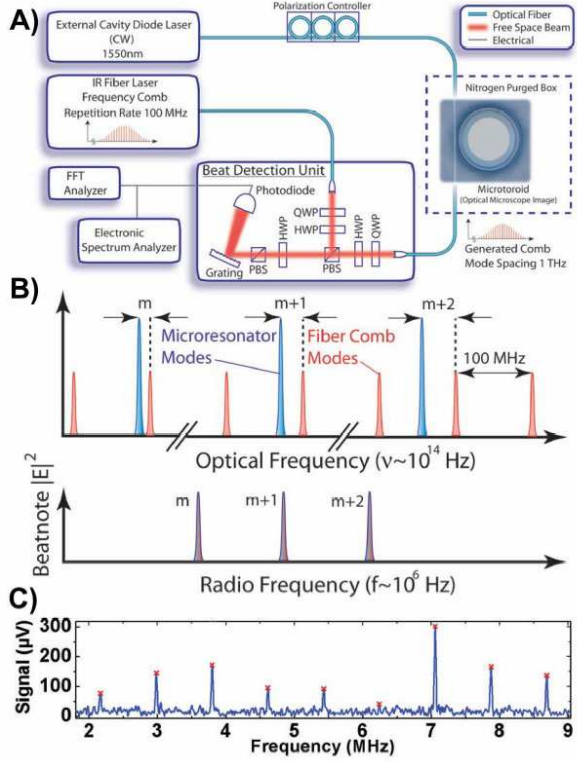}
\end{center}
\caption{Parametric beat note setup. Panel (A) shows the experimental setup
consisting of an external cavity laser (ECL) coupled to an ultra-high-Q
monolithic microresonator in a nitrogen environment via a tapered fiber. The
parametric output is coupled into one port of a beat note detection unit
(BDU). The second port of the BDU is coupled to a mode-locked femtosecond
erbium doped fiber laser that serves as a reference comb. A grating is used to
select a spectral region of the Kerr modes and a PIN Si photodiode records
their beatings with the reference comb (See supplementary information for
details). Panel (B) shows the measurement principle. The beating of the
reference comb with the parametric lines yields beat frequencies in the
radio-frequency domain. Panel (C): RF Spectrum of 9-simultaneously oscillating
Kerr modes, exhibiting a uniform spacing.}%
\end{figure}

\begin{figure}[tbh]
\begin{center}
\includegraphics[width=8cm]{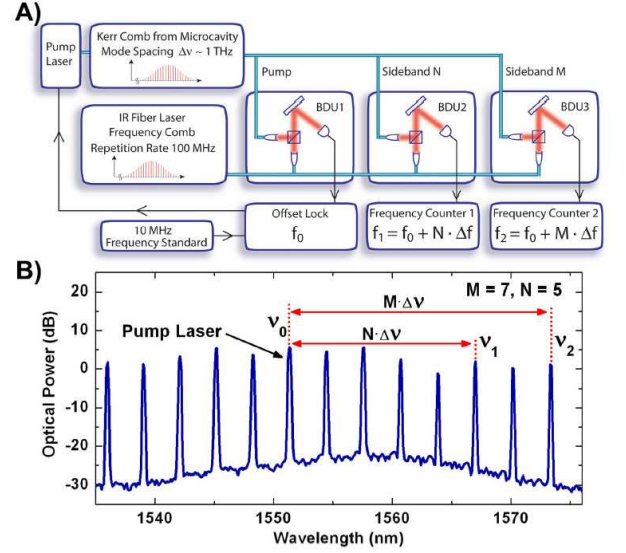}
\end{center}
\caption{{}Probing the equidistance of the comb structure. Panel (A):
Simplified schematic of the setup which consists of three beat note detection
units (BDUs) to measure the beating of three Kerr modes with the fiber based
reference comb. All BDUs are referenced to the MPQ hydrogen Maser as frequency
standard. The first BDU is used to implement a phase lock between one comb
line of the reference comb and the pump laser (which constitutes one mode of
the Kerr comb). Panel (B): The parametric spectrum under consideration for
validating the equidistance of the comb modes.}%
\end{figure}To verify that the Kerr comb indeed contains equidistant
frequencies, we employed a fiber-laser based optical frequency
comb\cite{Kubina2005} from Menlo Systems (termed \textquotedblleft reference
comb\textquotedblright\ in the remaining discussion) as a reference grid whose
repetition rate is $f_{\mathrm{rep}}=100$ MHz. The principle underlying our
measurement is that the beating generated on a photodiode by superimposing the
reference comb with the Kerr comb will produce beat notes which constitute a
unique replica of the optical spectrum in the radio frequency domain, provided
that the highest produced beat note in the detection process is
$<f_{\mathrm{rep}}/2$ (cf. Fig. 2b), similar to multi-heterodyne frequency
comb spectroscopy\cite{Schliesser2005}. Specifically, if the Kerr comb is
equidistant, the beat notes with the reference comb will constitute an
equidistant comb in the RF domain (with frequency spacing $\Delta f$, where
$\Delta f=\left(  \Delta\nu\ \mathrm{mod}\ f_{\mathrm{rep}}\right)  $). Fig.
2a shows the experimental setup for the optical beat measurement. In brief, an
external cavity laser at 1550 nm was used as pump laser (cf. Fig 2 main
panel). The Kerr lines of the microcavity were superimposed with the reference
comb in a beat note detection unit (BDU), consisting of polarizing optics to
combine the reference and Kerr combs and a grating to select the desired
region of spectral overlap. In this manner, the beating of 9 simultaneously
oscillating parametric modes (covering $>50$ nm of wavelength span) were
recorded, as shown in Fig. 2c. Remarkably, from the equidistant spacing of the
radio-frequencies, it is found that the generated sidebands are
\textit{equidistant} to within less than 5 kHz (as determined by the
measurement time of 200 $\mu$s). \newline

To improve the accuracy, we developed an additional experiment where we
measured the beat-notes of three Kerr modes with the fiber-reference comb
using three separate BDUs (cf. Fig. 3a) each counting a single beat of radio
frequency ($f_{0}$, $f_{1}$, $f_{2}$). A signal-to-noise ratio exceeding 30 dB
in 500 kHz bandwidth was achieved, sufficient to use radio-frequency counters,
which were all referenced to a 10 MHz reference signal provided by the MPQ
hydrogen Maser. The beat-note measured on BDU1 was used to implement an offset
lock between a single reference comb line and the pumping laser by a known
offset frequency ($f_{0}$). The second (third) counter measured the
N$^{\mathrm{th}}$ (M$^{\mathrm{th}}$) mode of the Kerr comb as shown in Fig.
3. For equidistant mode spacing, the second (third) BDU gives rise to the beat
frequency $f_{1}=f_{0}+\Delta f\times N$ ($f_{2}=f_{0}+\Delta f\times M$). The
uniformity of the Kerr comb was then checked by comparing the variation in the
mode spacing, i.e. $\epsilon\equiv\frac{f_{2}-f_{1}}{M-N}-\frac{f_{1}-f_{0}%
}{N}$. Alternatively, direct counting of the ratio $\left(  f_{1}%
-f_{0}\right)  /\left(  f_{2}-f_{0}\right)  =\frac{N}{M}$ was implemented
(using frequeny mixing and ratio counting, cf. SI). Fig. 4b shows the result
of this measurement (for $N=5$, $M=7$) and a counter gate time ($\tau$) of 1
second and more than 3000 records. The scatter in the data follows a Gaussian
distribution (and follows a $\frac{1}{\sqrt{\tau}}$ dependence of the Allan
deviation, cf. Fig. 4b). The cavity modes of this measurement span over
approx. 21 nm and yield a \textit{deviation from the mean} of $\pm5$ mHz. Note
that the wavelength span that could be used for the measurement is currently
limited by the gain bandwidth of an EDFA, which had to be used to amplify the
reference comb to have sufficient power to run three BDUs simultaneously. The
results for different gate times and for the two different counting methods
are shown in Table 1 (the complete list is contained in the supplementary
information). The weighted average of these results verifies the uniformity of
the comb spacing to a level of $7.3\times10^{-18}$ (when referenced to the
optical carrier). Normalized to the bandwidth of the measured Kerr lines (2.1
THz), this corresponds to $5.2\times10^{-16}$. This accuracy is on par with
measurements for fiber based frequency combs\cite{Kubina2005} and confirms
that the generated Kerr combs exhibit uniform mode spacing.

\begin{table}[ptb]%
\begin{tabular}
[c]{|l|l|l|l|l|l|}\hline
$\tau(s)$ & N & Mean ($m\operatorname{Hz}$) & $\sigma(m\operatorname{Hz})$ &
$\epsilon$ & Technique\\\hline\hline
1 & 3493 & -0.9 $\pm$ 5.5 & 322 & 2.7$\times$10$^{-17}$ & 2 Counters\\\hline
3 & 173 & 5.8 $\pm$ 12.6 & 165 & 6.3$\times$10$^{-17}$ & Ratio\\\hline
10 & 22 & -17.9 $\pm$ 15.0 & 70 & 7.5$\times$10$^{-17}$ & Ratio\\\hline
30 & 39 & 1.7 $\pm$ 7.4 & 46 & 3.7$\times$10$^{-17}$ & Ratio\\\hline
100 & 42 & -0.3 $\pm$ 2.7 & 17 & 1.4$\times$10$^{-17}$ & Ratio\\\hline
300 & 14 & -0.8 $\pm$ 2.8 & 11 & 1.4$\times$10$^{-17}$ & Ratio\\\hline
\end{tabular}
\caption{Summary of the experimental results on the accuracy of the mode
spacing. The weighted mean of all measurements (including SI) yields a
relative accuracy of $7.3\times10^{-18}.$}%
\label{tab:counts}%
\end{table}

\begin{figure}[tbh]
\begin{center}
\includegraphics[width=8cm]{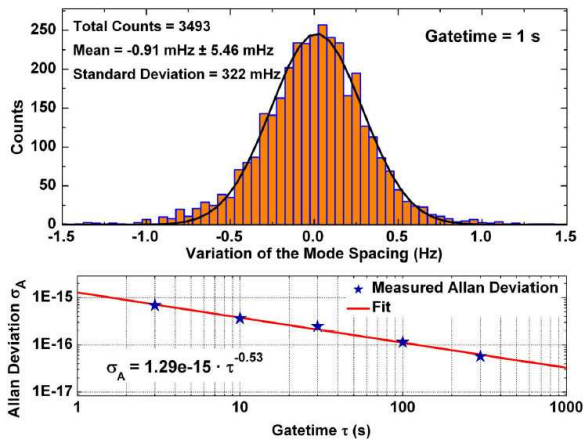}
\end{center}
\caption{{}Frequency counting experiment. Panel (A): The deviation from
equidistant mode spacing for a gate time of 1 second for the parametric
spectrum and measurement setup described in Fig 3. For this measurement 3493
points were recorded. The solid red line is a Gaussian fit to the
distribution. The deviation from the mean implies an accuracy of the mode
spacing at the mHz level, confirming the comb like structure of the parametric
spectrum. Panel (B): Allan deviation as a function of gate time, exhibiting an
inverse square-root dependence on the gate time, as determined by the fit
(solid red line).}%
\end{figure}Next we investigated the role of dispersion underlying the
observed comb generation. Dispersion in whispering-gallery-mode (WGM)
microcavities is characterized by the deviation in the free spectral range
$\Delta\nu_{\mathrm{FSR}}=\left(  \nu_{m+1}-\nu_{m}\right)  -\left(  \nu
_{m}-\nu_{m-1}\right)  =\nu_{m+1}-\nu_{m-1}-2\nu_{m}$ and has two
contributions. Geometrical dispersion accounts for a negative FSR dispersion,
given by $\Delta\nu_{\mathrm{FSR}}\approx-0.41\frac{c}{2\pi\cdot n\cdot
R}\cdot m^{-5/3}$ where $R$ the cavity radius (cf. supplementary information).
Material dispersion on the other hand is given by $\Delta\nu_{\mathrm{FSR}%
}\approx\frac{1}{4\pi^{2}}\frac{c^{2}\cdot\lambda^{2}}{n^{3}\cdot R^{2}}\cdot
GVD$, where $GVD=-\frac{\lambda}{c}\frac{\partial^{2}n}{\partial\lambda^{2}}$
is the group velocity dispersion parameter. Since the GVD of silica is
positive for wavelength greater than 1.3 $\mu$m (anomalous dispersion), it can
compensate the intrinsic resonator dispersion (causing $\Delta\nu
_{\mathrm{FSR}}>0$). Indeed we measured a positive dispersion (cf. SI) which
equates to only ca. 20 MHz over a span of ca. 60 nm. This low value indicates
that the present experiments are carried out close to the zero dispersion
wavelength, in agreement with theoretical predictions. \newline Note that the
residual cavity dispersion exceeding the \textquotedblleft
cold\textquotedblright\ cavity linewidth does not preclude the parametric comb
generation process. This can be explained in terms of a nonlinear optical mode
pulling process as reported in Ref. \cite{Kippenberg2004a}. The strong CW pump
laser will induce both self-phase modulation (SPM) and cross-phase modulation
(XPM)\cite{agrawal}, the latter being twice as large as the former. The
resultant XPM and SPM induced refractive index changes will shift the cavity
resonance frequencies by different amounts, thereby causing a net change in
the (driven) cavity dispersion from its \textit{passive} (un-driven)
value\cite{Kippenberg2004a}. This nonlinear mode pulling can provide a
mechanism to compensate the residual cavity dispersion. \newline Regarding
future experimental work in light of applications in metrology, we note that
absolute referencing can be attained by locking the pump laser to a known
atomic transition and locking the mode spacing to a microwave reference (such
as a Cs atomic clock). The latter requires that the two degrees of freedom of
the comb, its repetition rate (i.e. mode spacing, $\Delta\nu$) and frequency
offset, i.e. $\nu_{\mathrm{CEO}}=\left(  \nu_{0}\ \mathrm{mod}\ \Delta
\nu\right)  $ to be controlled independently. Indeed it could already been
shown in a proof of concept experiment that it is possible to lock two modes
of the Kerr comb simultaneously to two modes of the reference comb, showing
that we are able to control both $\nu_{\mathrm{CEO}}$ and $\Delta\nu$. The two
actuators used for this lock are the detuning of the pump laser from the
microcavity resonance and the pump power, which affects the optical pathlength
of the cavity via the thermal effect and the nonlinear phase shift. \newline
Pertaining to the implications of our work, we note that the present
observation of a monolithic frequency comb generator could potentially prove
useful for frequency metrology, given however further improvements. Evidently
a readily measurable repetition rate would prove useful when directly
referencing the optical field to a microwave signal\cite{Cundiff2003}. To this
end a 660-$\mu$m-diameter microcavity would already allow operating at
repetition rates $<100$ GHz, which can be detected using fast photodiodes. On
the other hand, a large mode spacing as demonstrated here could prove useful
in several applications, such as line-by-line pulse shaping, calibration of
astrophysical spectrometers or direct comb spectroscopy. The high repetition
rate from an on chip device may also prove useful for the generation of
multiple channels for high capacity telecommunications (spacing 160 GHz) and
for the generation of low noise microwave signals. Furthermore, we note that
parametric interactions do also occur in other types of microcavities - e.g.
CaF$_{2}$\cite{Savchenkov2004a} - provided the material exhibits a third order
nonlinearity and sufficiently long photon lifetimes. As such the cavity
geometry is not conceptually central to the work and the reported phenomena
should become equally observable in other types of high-Q microresonators,
such as silicon, SOI or crystalline based WGM-resonators. Indeed the recent
observation of net parametric gain\cite{Foster2006} on a silicon chip is a
promising step in this direction.

\bigskip

\textit{Acknowledgements:} The authors thank T. W. H\"{a}nsch, Th. Udem, K. J.
Vahala and Scott Diddams for critical discussions and suggestions. TJK
acknowledges support via an Independent Max Planck Junior Research Group. This
work was funded as part of a Marie Curie Excellence Grant
(MEXT-CT-2006-042842), the DFG funded Nanoscience Initiative Munich (NIM) and
a Marie Curie Reintegration Grant (MIRG-CT-2006-031141). The authors
gratefully acknowledge J. Kotthaus for access to clean-room facilities for
sample fabrication.

\bibliographystyle{unsrt}
\bibliography{pascal_mpq}

\newpage

\begin{widetext}
\section{Generation of Kerr Combs at lower repetition rates}
Figure \ref{fig:lowrepcomb} shows the Kerr comb spectrum at a lower
repetition rate mentioned in the main paper. The repetition rate is $375\,%
\mathrm{GHz}$, corresponding to a free spectral range of $3\,\mathrm{nm}$.
With larger samples it should be possible to generate repetition rates
smaller than $100\,\mathrm{GHz}$ which permits the direct measurement of the
repetition rate with high-bandwidth photodiodes.
\begin{figure}[hbt!]
\begin{center}
\includegraphics[width=0.7\textwidth]{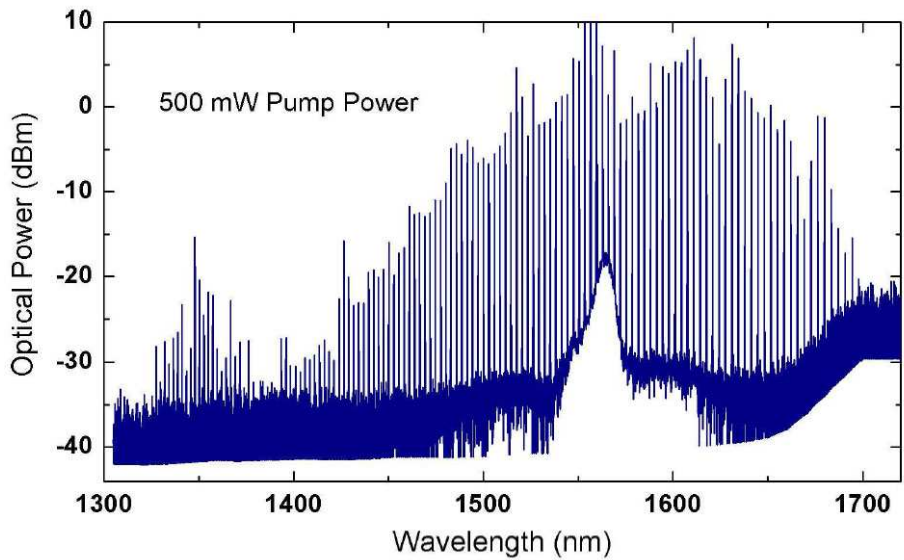}
\end{center}
\caption{Kerr comb generated in an $177 \mbox{-} \protect\mu \mathrm{m}$-diameter
toroid. The total power in the spectrum (pump line + generated sidebands) is around $500\
\mathrm{mW}$ distributed over more than $134$ lines. The free spectral range
is $3\ \mathrm{nm}$.}
\label{fig:lowrepcomb}
\end{figure}
\section{Beat note experiments between the Fiber Laser Comb and Kerr Comb}
To demonstrate the equidistant nature of the parametric Kerr lines,
a reference frequency comb in the form of a mode locked erbium fiber
laser is used (from Menlo Systems GmbH). The principle underlying
the measurement is similar to the concept of multi-heterodyne
spectroscopy\cite{Schliesser2005}. Assuming that the
reference comb produces a spectrum with frequencies $f_\mathrm{ceo}+n\cdot f_{\mathrm{%
rep}}$ (where $f_{\mathrm{rep}}$ is the repetition rate,
$f_\mathrm{ceo}$ is the carrier envelope offset frequency and $n$ is an integer number of order $%
2\cdot 10^{6}$) and the Kerr comb produces frequencies
$\nu_{0}+m\cdot \Delta\nu$ ($m$ integer), the signal generated by
interfering the two combs will have an imprinted radio frequency
(RF) beat note spectrum. If the reference comb's repetition rate is
adjusted such that a multiple of it is close to the Kerr mode
spacing, i.e. $m_{0}\cdot f_{\mathrm{rep}} \approx \Delta\nu$ (with an integer $m_{0}$%
), then the $N$ different Kerr comb lines will generate $N$
different RF beat notes which will again be evenly spaced, i.e.
their frequencies are $f_{0}+\Delta f \cdot k$ (with $\Delta f
=(\Delta \nu \, \mathrm{mod}\, f_{\mathrm{rep}})$ and $k$ integer).
\newline
The experimental setup is depicted in Figure 2 of the main paper. A
tunable external cavity diode laser (ECDL) is used to pump a
microtoroid resonance as detailed in \cite{Kippenberg2004b} and
\cite{Kippenberg2004a}. Since the cavity resonances are polarization
dependent, a in-fiber polarization controller is used to adjust the
polarization of the pump laser. The microtoroid is placed in a
sealed enclosure containing a nitrogen atmosphere, to avoid the
deposition of water on the surface of the silica toroid which has
strong absorption bands in the $1550\mbox{-nm}$ regime. In the
microresonator a spectrum of modes is generated via nonlinear
parametric interactions and four-wave mixing (see main paper). The
output signal of the tapered optical fiber (containing the
parametric modes that are outcoupled from the microresonator back to
the tapered fiber) is split by two $3\ \mathrm{dB}$ couplers and
monitored with a photodiode connected to an oscilloscope and an
optical spectrum analyzer. Another fraction of the taper output is
sent to a \textquotedblleft beat detection unit\textquotedblright\
(BDU) and superimposed with a fiber-laser-based reference frequency
comb with a repetition rate of $100\ \mathrm{MHz}$\cite{Kubina2005}.
The BDU consists of quarter wave plates and half wave plates to
prepare orthogonal linear polarization in the two input beams, which
are subsequently combined using a polarizing beam splitter. By means
of a half-wave plate, an adjustable linear combination of the two
input beams' polarizations is then rotated onto the transmission
axis of a polarizer, where the two input beams interfere. To
increase the signal-to-noise ratio (SNR), the spectral region
containing the Kerr comb lines is selected by a grating and finally
detected with a PIN InGaAs photodiode (Menlo Systems FPD 510). An
oscilloscope with a built-in FFT routine is utilized to analyze the
radio frequency spectrum. For rough analysis an electronic spectrum
analyzer is used. Since the repetition rate of the reference comb is
around $100\ \mathrm{MHz}$ the beat note frequencies between a laser
line and the reference comb are in the range of $0\ \mathrm{MHz}$ to
$50\ \mathrm{MHz}$. Now the repetition rate of the reference comb is
adjusted until $(\Delta \nu \, \mathrm{mod} \, f_{\mathrm{rep}})$ is a
small frequency such that for all $k$ of interest the condition $0<
f_0 + k\cdot \Delta f < f_\mathrm{rep}/2$ is fulfilled. The
observation of an equidistant RF beat \textquotedblleft
comb\textquotedblright\ then provides proof for the equidistance of
the Kerr comb. \
\section{Measuring the accuracy of the mode spacing using heterodyne
spectroscopy}
\subsection{Measuring with two counters}
To verify the equidistance of the Kerr comb modes it is necessary to
know the frequencies of three Kerr comb modes simultaneously. The
frequency counting is achieved by using radio frequency counters
that are connected to a photodiode in a beat note detection unit
(cf. figure 2 in the main paper). To determine the frequencies of
three Kerr comb modes at the same time, three beat note detection
units (BDU) have been built. By tuning the grating of the BDUs it is
possible to measure the beat note frequency of a single Kerr comb
line with a reference comb line. For simplicity reasons, one BDU is
used to lock the diode laser pumping the microcavity to a single
mode of the reference comb. Additionally the repetition rate of the
reference comb is locked to a frequency of around 100 MHz,
stabilized by a 10 MHz frequency standard generated by an in-house
hydrogen maser. The two remaining beat detection units are placed at
the output of the microcavity and the gratings are adjusted in a way
that each of them counts a different Kerr comb mode. Note that the
output of the reference comb had to be amplified with an EDFA to
obtain sufficient power to run three BDUs simultaneously (a single
line of the reference comb contains ca. 10 nW optical power). With
this setup it was possible to achieve signal-to-noise ratios for the
Kerr sideband beat notes of more than 30 dB at a resolution
bandwidth of 500 kHz (Additional RF filters with a 3-dB-bandwidth of
3 MHz have been used to filter out background noise). In the present
experiment we focused on counting the 5$^{th}$ (beat note frequency
$f_1$) and the 7$^{th}$ Kerr comb sideband (beat note frequency
$f_2$), whereas the pump laser was phase locked to the fiber laser
reference comb such that its beat with the reference comb was fixed
to a frequency $f_0$. Note that the pump laser already constitutes
one tooth of the Kerr comb. For an equally spaced Kerr comb we
therefore expect $f_1=f_0 + N \cdot \Delta f$ and $f_2=f_0+ M \cdot
\Delta f$ with $N=5$ and $M=7$ to be the beat note frequencies of
the sidebands. The variation of the mode spacing $\epsilon$ of the
Kerr comb is given by
\begin{equation}
\epsilon=\frac{f_2-f_1}{M-N} - \frac{f_1-f_0}{N} \ \ \ \mathrm{,}
\label{eqn:variation}
\end{equation}
which is zero for an equally spaced comb. With the measured values
for $f_1$ and $f_2$ and the known frequency $f_0$ it is possible to
calculate the variation of the mode spacing $\epsilon$. The two
counters are referenced to the same frequency standard as the offset
lock for the pump laser and are externally triggered with a signal
from a pulse generator. This external triggering was necessary since
the mode spacing of the Kerr comb was fluctuating by approximately
40 kHz r.m.s., giving rise to ``breathing'' of the Kerr comb
modes(cf. figure \ref{fig:driftFSR}). Hence, it proved cricital for
a high accuracy that the two counters measured simultaneously, to
allow the cancellation of the common fluctuations.
\begin{figure}[hbt!]
\begin{center}
\includegraphics[width=0.6\textwidth]{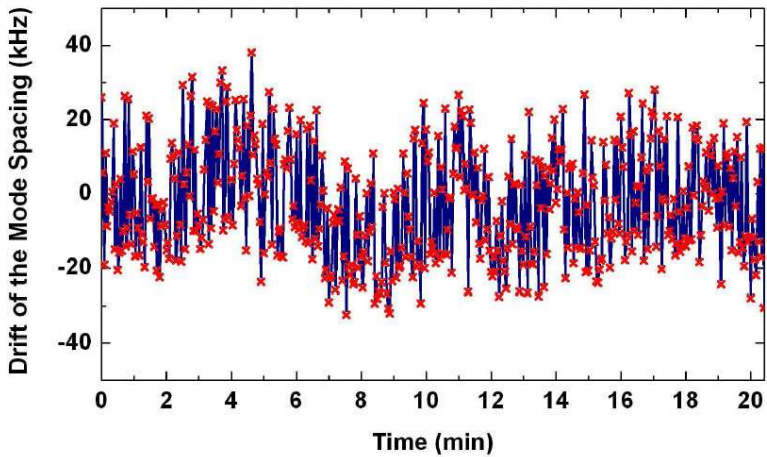}
\end{center}
\caption{Drift of the mode spacing of the Kerr comb when not stabilized. The mode spacing exhibits fluctuations of approximately 40 kHz for short time scales and some slower thermal drifts for time scales of several minutes. Note that these fluctuations of the mode spacing do not affect the equidistance of the modes of the Kerr comb since they are fluctuating simultaneously.}
\label{fig:driftFSR}
\end{figure}
\subsection{Measuring the ratio of the distance to the sidebands}
To avoid the synchronization problems mentioned before, the
experimental setup depicted in figure \ref{fig:setupratio} was used.
In brief, the three counter signals were first electronically mixed
with $f_0$ and filtered yielding only the distance between pump and
the N$^\mathrm{th}$ (M$^\mathrm{th}$) Kerr-sidebands. With this
setup, a slightly smaller standard deviation of the measurements
could be achieved by using just one counter with two inputs to
measure the ratio $R$ of the distance between the pump beat and the
two sideband beats,
\begin{equation}
R = \frac{f_2 - f_0}{f_1 - f_0}.
\end{equation}
Solving this for $f_2$ and using equation \ref{eqn:variation} we
obtain the dependence of the variation of the mode spacing
$\epsilon$ from the ratio R:
\begin{equation}
\epsilon = \frac{f_1 - f_0}{M-N} \cdot R + \left( f_0 - f_1 \right) \cdot \left( \frac{1}{M-N} + \frac{1}{N}  \right)
\end{equation}
Using the frequency difference $f_1 - f_0$, which was set to
approximately 10 MHz, it is possible to derive the variation of the
mode spacing $\epsilon$ by measuring the frequency ratio R.
\begin{figure}[hbt!]
\begin{center}
\includegraphics[width=0.6\textwidth]{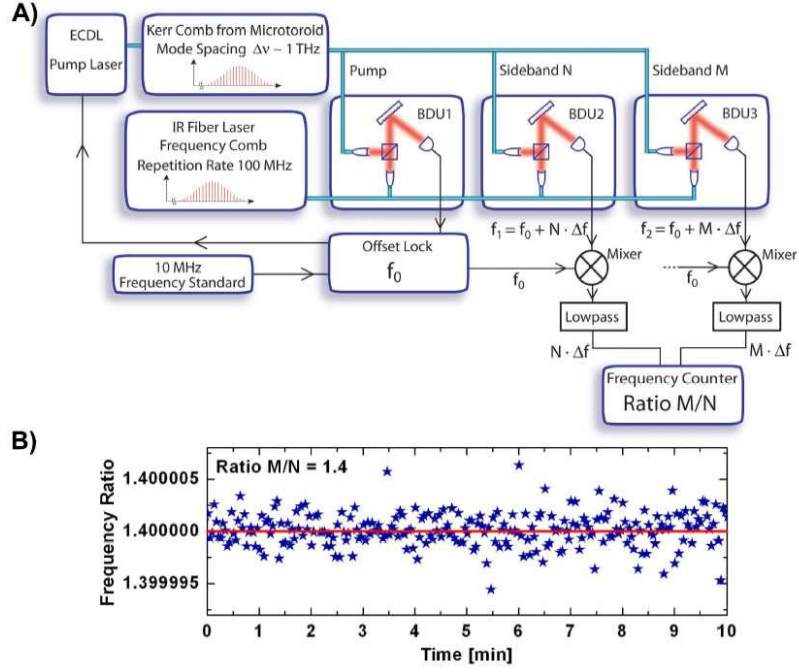}
\end{center}
\caption{Panel A). Experimental setup to measure the ratio of the
frequency separation between pump laser and two different Kerr comb
sidebands. ECDL = External Cavity Diode Laser. Beat note detection
unit 1 (BDU1) is used to phase lock the pump laser line from the
Kerr comb to the reference comb with an offset frequency $f_0$. BDU2
(BDU3) is adjusted to measure the beat note frequency between the
N$^{th}$ (M$^{th}$) Kerr comb line and the reference comb. By mixing
these frequencies down with the offset lock frequency $f_0$ using
electronic mixers, new frequencies $N \cdot \Delta f$ and $M \cdot
\Delta f$ are generated. The ratio of these frequencies is $M/N=1.4$
for the 7$^{th}$ and the 5$^{th}$ sideband.  Panel B) shows a
measurement of the frequency ratio of the radio frequency beat notes
generated from the 7$^{th}$ and the 5$^{th}$ sideband of the Kerr
comb.} \label{fig:setupratio}
\end{figure}
\section{Experimental Results of the Counter Measurements}
Table \ref{tab:counts} shows the experimental results from the
measurements of the Kerr comb equidistance. Note that a total of 9
data points out of the 8382 measurements from table \ref{tab:counts}
have been removed from analysis. These data points have been
separated by the other data points of the respective measurements by
more than 15 standard deviations. Assuming a Gaussian distribution
(which was indeed found for the remaining 8373 measurements) the
probability of measuring a point 15 standard deviations off as given
by the cumulative error function is $(1-\mathrm{erf(15/\sqrt{2}}))
\approx 7.3 \times 10^{-51}$. These points are believed to originate
from some local perturbations in the lab leading to a temporary
reduction of the signal-to-noise level of the radio frequency beat
notes. The weighted mean $\bar{\epsilon}_w$ in table
\ref{tab:counts} has been calculated with the inverse squared
standard error of the mean as weight:
\begin{equation}
\bar{\epsilon}_w = \frac{\sum \epsilon / \sigma_{\epsilon}^2}{\sum 1 / \sigma_{\epsilon}^2}
\end{equation}
\begin{equation}
\sigma_{\epsilon_w}^2 = \frac{1}{\sum 1 / \sigma_{\epsilon}^2}
\end{equation}
The weighted mean calculated from all measurements leads to a variation of the modespacing of $\bar{\epsilon}_w = -0.8 \ \mathrm{mHz} \pm 1.4 \ \mathrm{mHz}$. Normalized to the optical carrier frequency of $192$ THz, this leads to an accuracy of the equidistance of $7.3 \times 10^{-18}$.
\begin{table}[hbt!]
\centering
\begin{tabular}{|l|l|l|l|l|l|}
\hline
\textbf{Gate time (s)} & \textbf{Readings} & \textbf{Mean Value for $\epsilon$ (mHz)}& \textbf{StdDev of $\epsilon$ (Hz)}& \textbf{Counting Method} \\ \hline \hline
%0.003& 283&    -6060 $\pm$ 4780 &  80& ratio    \\ \hline
%0.01&  316&    3165 $\pm$ 1567&    28& ratio    \\ \hline
0.03&   217&    -33 $\pm$ 556 & 8.2&    ratio    \\ \hline
0.1&    223&    -80 $\pm$ 181&  2.7&    ratio    \\ \hline
0.3&    293&    2.4 $\pm$ 50.1& 0.86&   ratio    \\ \hline
1&  3493&   -0.91 $\pm$ 5.46&   0.32&   2 counters   \\ \hline
1&  3499&   3.9 $\pm$ 10.1& 0.60&   2 counters   \\ \hline
1&  98& -40.1 $\pm$ 27.4&   0.27&   ratio    \\ \hline
1&  179&    8.0 $\pm$ 25.5& 0.34&   ratio    \\ \hline
3&  173&    5.8 $\pm$ 12.6& 0.17&   ratio    \\ \hline
10& 22& -17.9 $\pm$ 15.0&   0.070&  ratio    \\ \hline
30& 39& 1.65 $\pm$ 7.41&    0.046&  ratio    \\ \hline
60& 72& -1.88 $\pm$ 3.00&   0.025&  ratio    \\ \hline
100&    18& 1.12 $\pm$ 5.98&    0.024&  ratio    \\ \hline
100&    42& -0.26 $\pm$ 2.69&   0.017&  ratio    \\ \hline
300&    14& -0.82 $\pm$ 2.83&   0.011&  ratio    \\ \hline \hline
\multicolumn{2}{|r|}{Weighted Mean $\bar{\epsilon}_w$: } &-0.8 mHz $\pm$ 1.4 mHz &  - & - \\ \hline
\end{tabular}
\caption{Complete list of the Kerr comb measurements with frequency counters. StdDev = standard deviation of the distribution. The last column shows the used method to acquire the data: ``2 counters'' stands for the measurements with two externally triggered counters (one for each Kerr sideband) and ``ratio'' stands for the method with one counter that directly measures the ratio between the distance between pump and two different Kerr comb lines (both methods are explained in the preceding section). As expected, the standard deviation of the measurements reduces with increasing gatetime. The total measurement time is 6 h 37 min.}
\label{tab:counts}
\end{table}
\section{Measurement of Cavity Dispersion}
To measure cavity dispersion, we employ the arrangement shown in figure \ref%
{fig:setupdispersion}. In brief, we first lock an external cavity laser
around $1550 \ \mathrm{nm}$ to one of the fundamental WGM cavity modes (the
same resonance that gives rise to cascaded sidebands at higher power). The
cavity resonance of the monolithic microresonator is locked to the external
cavity laser by virtue of the thermal self locking technique\cite{Carmon2004}%
. The power is chosen to be far below the parametric threshold $<85
\ \mu \mathrm{W}$ but sufficient to entail a stable lock. Next, the
frequency comb is offset-locked to the external cavity laser by
recording the beat note signal in a separate beat note detection
unit (for working principle of the beat detection unit see last
section). To achieve stable locking the generated beating is
filtered and amplified yielding a SNR of ca. $25-30 \ \mathrm{dB}$
(at a resolution bandwidth of $400 \ \mathrm{kHz}$). For dispersion
measurement the frequency comb must be locked at an arbitrary
detuning with respect to the ECDL. The latter is accomplished by
mixing the beat note with a (variable) reference signal
($f_{\mathrm{offset}}$) down to $10$ $\,\mathrm{MHz}$ and
implementing a phase lock with feedback on the fiber comb's
repetition rate ($f_\mathrm{rep}$) by controlling the cavity length
using a mirror mounted on a piezoelectric tube (Note that all RF
generators and analyzers are stabilized using an in-house
10-MHz-reference). Owing to the fact that the cavity linewidth is
$<5$
$\,\mathrm{MHz}$ and the repetition rate of the fiber comb (FC) is $100$ $\,\mathrm{MHz}$%
, not more than $one$ FC comb mode at a time can be resonant with one
microresonator mode. Since measuring the coupling of an individual comb
mode into the resonator in transmission is difficult, we measure the
reflection of the cavity induced by modal coupling\cite{Kippenberg2002}. By
variation of $f_{\mathrm{offset}}$\ (and by recording simultaneously $f_{%
\mathrm{rep}}$) this allows to resolve the linewidth of individual cavity
modes in reflection when using an OSA in zero-span mode. Hence this measurement provides an accurate means to
measure frequency gap (free spectral range) between two cavity resonances $%
\nu_{m}$ and $\nu_{m + \Delta m}$ modulo the repetition rate of the
fiber comb $((\nu _{m}-\nu _{m+\Delta m} )
\mathrm{mod}f_\mathrm{rep})$. The low power of the individual FC lines
(ca. $10 \ \mbox{nW}$) ensures that the\ probed cavity mode is not
thermally distorted. To remove the ambiguity in the number of comb
lines $(n)$ between the FSR of the cavity i.e. $n=\lfloor (\nu
_{m}-\nu _{m + \Delta m})/f_{\mathrm{rep}}\rfloor $ a second
measurement was carried out with a different repetition rate, which
allowed to retrieve $n$. So the actual free spectral range between
two cavity resonances can be derived by:
\begin{equation*}
\nu_\mathrm{FSR} = f_{\mathrm{beatnote}} + n \cdot f_{\mathrm{rep}}
\end{equation*}
\begin{figure}[hbt!]
\begin{center}
\includegraphics[width=0.9\textwidth]{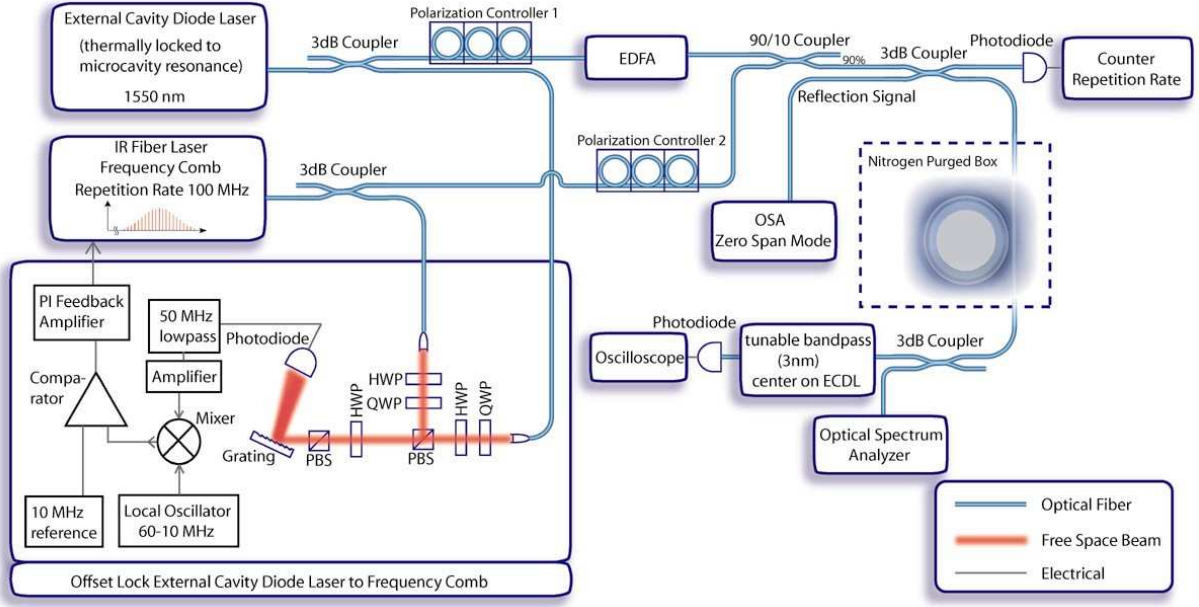}
\end{center}
\caption{Experimental setup of the dispersion measurement. The beat
detection unit on the lower left side is used to establish an offset
lock between the external cavity diode laser (ECDL) and the fiber
laser frequency comb. Therefore the signal from the photodiode in
the beat detection unit is first filtered with a $50\ \mathrm{MHz}$
lowpass to remove the strong signal of the $100\ \mathrm{MHz}$
repetition rate of the fiber laser comb. Subsequently the beat note
signal is mixed down to $10\ \mathrm{MHz}$ with a variable frequency
generator ($10..60\ \mathrm{MHz}$) and compared with a stable $10\
\mathrm{MHz}$ RF reference. The output of the comparator is sent to
a PI feedback amplifier which is connected to a piezo-mechanical
control of the repetition rate of the fiber laser. By adjusting the
variable frequency generator one can change the distance between the
laserline of the ECDL and the next comb line to an arbitrary value
between $0\ \mathrm{MHz}$ and $f_\mathrm{rep}/2$. The ECDL and the
fiber comb are furthermore coupled to the microcavity with a
microtoroid resonance thermally locked to the ECDL. To measure the
distance between two cavity resonances an optical spectrum analyzer
(OSA) in zero span mode is set to a wavelength of a different cavity
resonance than the one pumped by the ECDL. Next, the offset lock is
changed until a reflection signal of the fiber comb is detected on
the OSA. Once this is achieved the ECDL and one mode of the fiber
comb are on resonances with two different modes of the microcavity.
This means the FSR can be derived as $f_{\mathrm{beatnote}}+n\cdot
f_{\mathrm{rep}}$.} \label{fig:setupdispersion}
\end{figure}
Figure \ref{fig:dispersionmeasured} shows the experimental result of
the dispersion measurement. The used cavity had a free spectral
range (FSR) of $7.9$ nm, which corresponds to $0.96$ THz. Plotted in
figure \ref{fig:dispersionmeasured} is the accumulated dispersion of
the FSR, which we express for convenience as $ \left( \nu_{m+1} -
\nu_m \right) - \left( \nu_{1} - \nu_{0} \right)$. Here, the $\nu_m$
are the resonance frequencies of a ``cold'' microcavity. For this
measurement, $\nu_0$ is a resonance at 1585 nm ($189$ THz). From the
graph it can be derived that the accumulated dispersion is 2.6 MHz
per FSR (i.e. positive dispersion).
\begin{figure}[hbt!]
\begin{center}
\includegraphics[width=0.5\textwidth]{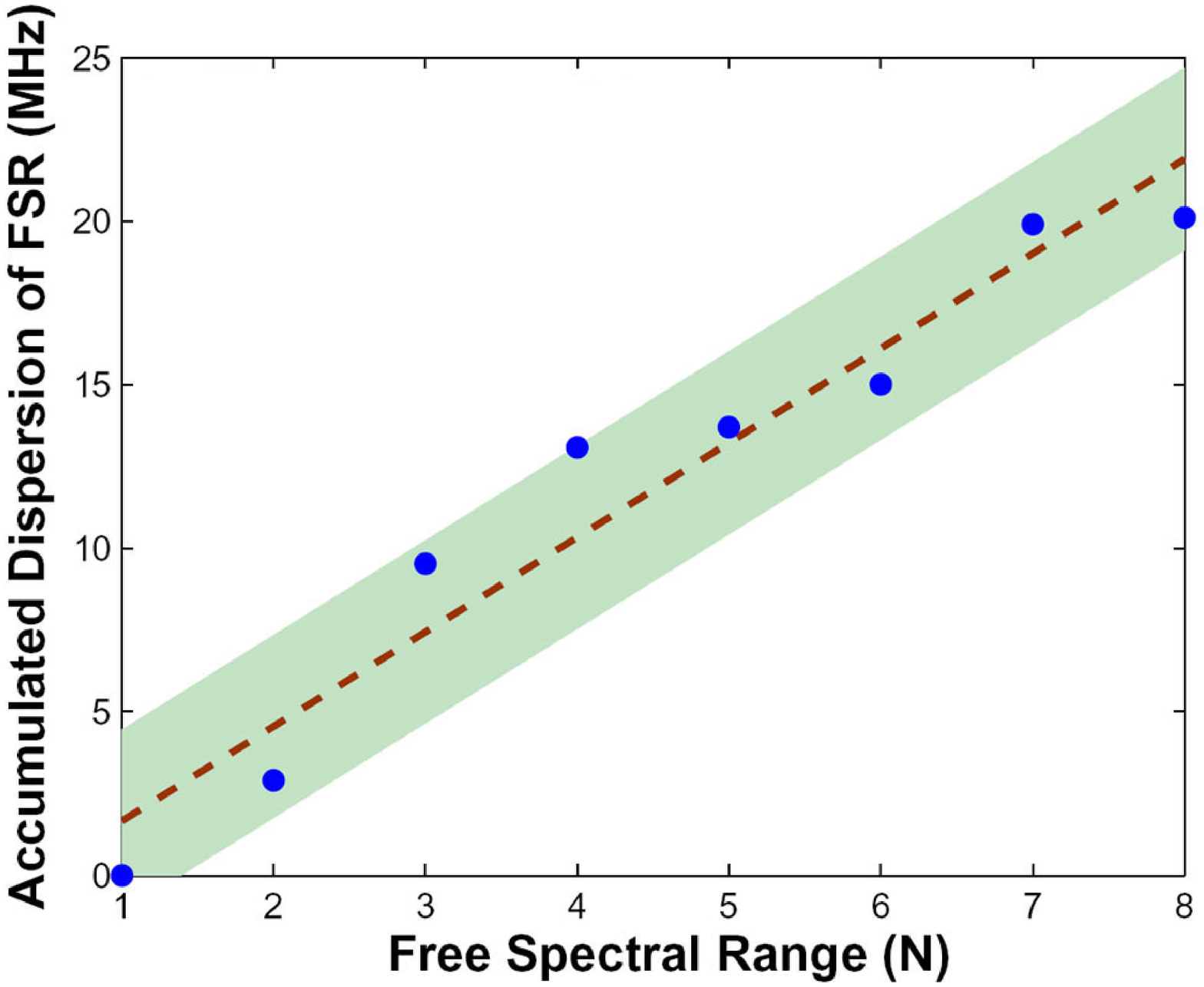}
\end{center}
\caption{Dispersion measurement of an 80-$\mu$m-diameter monolithic
microresonator. The figure shows the accumulated variation (i.e.
dispersion) of the free spectral range i.e. $\left( \nu_{m+1} -
\nu_m \right) - \left( \nu_{1} - \nu_{0} \right)$. The variation of
the FSR at higher frequencies (shorter wavelength) is referenced to
the free spectral range recorded between 1577 nm ($\nu_1$) and 1584
nm ($\nu_0$). The shaded region denotes experimental uncertainty,
the dotted line denotes a linear fit. As expected for a
whispering-gallery mode dominated by material dispersion, the free
spectral range increases for shorter wavelength.}
\label{fig:dispersionmeasured}
\end{figure}
\section{Dispersion Predictions}
The dispersion in our microcavities has two contributions. First,
whispering-gallery mode microcavities exhibit an intrinsic variation
of the free spectral range owing to the resonator geometry. The
resonance frequency of the fundamental mode of a microsphere is
approximately given by \cite{Schiller1993}
\begin{equation}
\nu_m=\frac{c}{2 \pi n R} \left( m+1/2+ \eta_1
\left(\frac{m+1/2}{2}\right)^{1/3}+\ldots
\right),
\end{equation}
where $c$ is vacuum light speed, $n$ the refractive index, $R$ the
cavity radius and $\eta_1$ the first zero of the Airy function
($\eta_1\approx 2.34$). Hence, the variation of the free spectral
range
\begin{equation}
\Delta\nu_\mathrm{FSR}=(\nu_{m+1}-\nu_m)-(\nu_{m}-\nu_{m-1})=
\nu_{m+1}+\nu_{m-1}-2\nu_m\approx \frac{\partial^2
\nu_{m}}{\partial m^2}
\end{equation}
is given by
\begin{equation}
\Delta\nu_\mathrm{FSR}= - \frac{c}{2 \pi n
R}\cdot\frac{\eta_1}{18}\left(\frac{m+1/2}{2}\right)^{-5/3}\approx -0.41\frac{c}{2 \pi n
R}m^{-5/3}<0
\end{equation}
Evidently, the free spectral range \emph{reduces} with increasing
frequency corresponding to a negative group velocity dispersion
(GVD), i.~e.\ low frequency modes exhibit a shorter round trip time
than high frequency modes. Supplementary figure S3 shows the
variation for a 40- and 80-micron-radius microsphere. \\
A second contribution comes from the dispersion of the fused silica
material constituting the resonator. Its contribution can be
estimated by considering that the refractive index $n$ is actually a
function of frequency (and therefore mode number $m$), $n\equiv
n(m)$. Neglecting geometric dispersion, the GVD of fused silica
alone would lead to a FSR variation of
\begin{equation}
\Delta\nu_\mathrm{FSR}\approx
\frac{\partial^2}{\partial m^2}\left( \frac{c}{2 \pi n(m) R}\cdot m\right)\approx
\frac{c^2 \lambda^2}{4 \pi^2 n^3 R^2}\cdot\mathrm{GVD},
\end{equation}
where
\begin{equation}
\mathrm{GVD}=-\frac{\lambda}{c}\frac{\partial^2 n}{\partial
\lambda^2}
\end{equation}
is the group-velocity dispersion of fused silica. This material
parameter is well-known to change its sign in the 1300-nm wavelength
region from about $-100\,\mathrm{ps/nm\cdot km}$ at 800 nm to
$+20\,\mathrm{ps/nm\cdot km}$ at 1550 nm. Combining the two
contributions, the positive sign of the GVD allows us in particular
to cancel the geometric dispersion of our resonators to some extent,
rendering the FSR nearly constant over a wide frequency span. Figure
\ref{fig:dispcalc} displays the FSR variation for an 80- and 160-micrometer diameter microsphere,
considering both material and geometric dispersion. Importantly, a
zero dispersion point close to our operating wavelength occurs. Note
that for a toroidal microcavity the location of the zero dispersion
point is expected to be shifted to shorter wavelengths owing to the
different resonator geometry. This expectation is borne out of
finite element simulations showing that the resonance wavelength for
a given $m$ value is shorter in a microtoroid cavity as compared to
a microsphere \cite{Kippenberg2004e}.
\begin{figure}[hbt!]
\begin{center}
\includegraphics[width=.6 \linewidth]{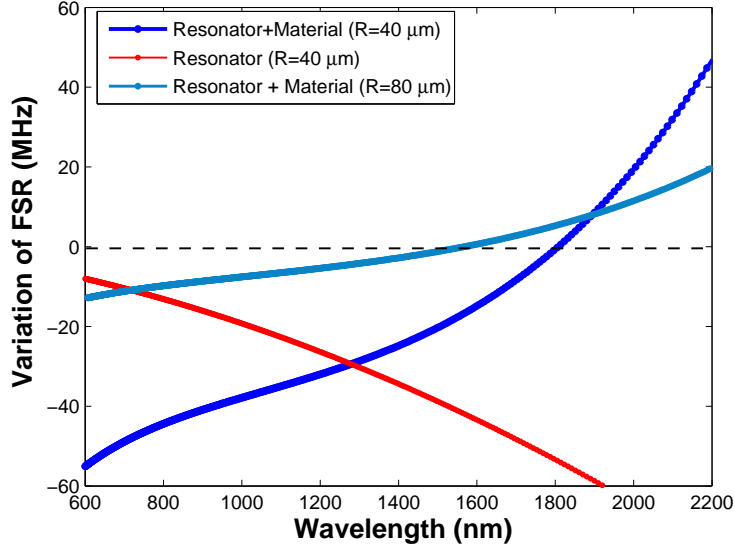}
\end{center}
\caption{Variation of the free spectral range of a
whispering-gallery microsphere resonator (i.e.
$\Delta\nu_\mathrm{FSR}=\nu_{m+1}+\nu_{m-1}-2\nu_m$). Shown is the
FSR\ dispersion for two resonator radii (40 $\protect\mu m$ and 80
$\protect\mu m$) including the effect of silica dispersion via the
Sellmeier equation. Resonance locations were calculated using an
asymptotic expansion of the microsphere resonance locations. Owing
to the different signs of silica material and resonator dispersion,
a zero dispersion point exists in the infrared.}
\label{fig:dispcalc}
\end{figure}
\section*{APPENDIX A: SYMBOLS USED THROUGHOUT THIS WORK}
\begin{tabular}
[c]{ll}%
\textbf{Symbols} & \textbf{Designation}\\
$\nu_{m}$ & Optical microcavity mode (with angular mode number$m$)\\
$\nu_\mathrm{FSR}$ & Optical microcavity\ free spectral range
$\left( \nu_\mathrm{FSR}=\left\vert \nu_{m}-\nu_{m+1}\right\vert \right)  $\\
$\Delta\nu_\mathrm{FSR}$ & Optical microcavity\ variation of the
free spectral range
$\left(  \Delta\nu_\mathrm{FSR}=\nu_{m+1}+\nu_{m-1}-2\nu_{m}\right)  $\\
$\nu_\mathrm{ceo}$ & Kerr comb carrier envelope offset frequency  \\
$\Delta\nu$ & Kerr comb mode spacing \\
$f_\mathrm{rep}$ & Fiber reference comb repetition rate\\
$f_\mathrm{ceo}$ & Fiber reference comb carrier envelope frequency\\
$f_{0,1,2}$ & Beat note unit (BDU) frequencies \\
$\Delta f$ & Frequency spacing of the multi-heterodyne beat comb
\end{tabular}
\end{widetext}

\end{document}